# Constructed measures and causal inference: towards a new model of measurement for psychosocial constructs

Tyler J. VanderWeele
Harvard University

**Abstract**. Psychosocial constructs can only be assessed indirectly, and measures are typically formed by a combination of indicators that are thought to relate to the construct. Reflective and formative measurement models offer different conceptualizations of the relation between the indicators and what is sometimes conceived of as a univariate latent variable supposed to correspond in some way to the construct. It is argued that the empirical implications of reflective and formative models will often be violated by data since the causally relevant constituents will generally be multivariate, not univariate. These empirical implications can be formally tested but factor analysis is not adequate to do so. It is argued that formative models misconstrue the relationship between the constructed measures and the underlying reality by which causal processes operate, but that reflective models misconstrue the nature of the underlying reality itself by typically presuming that the constituents of it that are causally efficacious are unidimensional. The ensuing problems arising from these misconstruals are discussed. A causal interpretation is proposed of associations between constructed measures and various outcomes that is applicable to both reflective and formative models and is applicable even if the usual assumptions of these models are violated. An outline for a new model of the process of measure construction is put forward. Discussion is given to the practical implications of these observations and proposals for the provision of definitions, the selection of items, item-by-item analyses, the construction of measures, and the interpretation of the associations of these measures with subsequent outcomes.

## Introduction

The model that dominates classical approaches to measurement and scale development, sometimes referred to as a reflective model, presupposes an underlying univariate latent variable that gives rise to measured indicators.[1-4] The latent variable is typically thought to correspond to some psychosocial construct of interest. It is often assumed that the underlying latent variable has causal efficacy,[4,5] even though all we observe are its indicators.

In this paper, I will present some empirical data and theoretical considerations that challenges whether either reflective or alternative so-called formative models are adequate. This will be facilitated by reviewing and deploying recently developed theory for causal inference for multiple version of treatment (MVT)[6,7] to develop alternative interpretations of exposure-outcome associations when the exposure used is a scale or index. I will show that the proposed interpretation under MVT theory holds for both reflective and formative models, and holds more generally still.[8-13] I will present an alternative model concerning the relationships between constructs, indicators, measures, and the true underlying constituents of reality, and I will discuss the practical implications of this for the process of measure construction.

## Causal Inference Under Multiple Versions of Treatment

The theory for causal inference under multiple versions of treatment[6,7] was originally developed to aid interpretation in settings wherein there was no unambiguous intervention to manipulate an exposure. In such settings, because different manipulations to shift the exposure might result in different effects on an outcome, the counterfactuals or potential outcomes[14-17] are not well-defined, and thus there is no single quantitative effect of *the* exposure.[6,7,18,19] It will be argued below that these issues are relevant for most psychosocial phenomena.

Consider first settings with an unambiguous exposure-intervention. Let A denote the exposure, Y an outcome, and C a set of pre-exposure covariates. Let $Y_a$ denote the potential outcome for Y if exposure A had been set to value a. The causal effect of a binary exposure A on outcome Y is defined, for an individual, by $Y_1 - Y_0$, and for the population by $E[Y_1 - Y_0]$. If the exposure is categorical or continuous, the values 1 and 0 can be replaced by arbitrary values, a and a*, respectively. We say that the effect of A on Y is unconfounded given C if $Y_a$ is independent of A conditional on C i.e. if, conditional on C, those with and without the exposure are comparable in their potential outcomes. If this is so and the technical consistency assumption holds, that when A=a then Y=$Y_a$, then we have[14-17]:

$$E[Y_1 - Y_0] = \sum_c \{E[Y|A = 1, c] - E[Y|A = 0, c]\} P(c).$$

The causal effect can thus be obtained by standardizing conditional observed outcome differences across exposure groups by the proportion in each stratum of C. In practice, this is often obtained by regressing Y on (A,C):

$$E[Y|a, c] = \beta_0 + \beta_1 a + \beta_2' c$$

Provided the regression model is correctly specified, the causal effect is then given by: $E[Y_1 - Y_0] = \sum_c \{E[Y|A = 1, c] - E[Y|A = 0, c]\} P(c) = \beta_1$.

Now consider the setting wherein there is not a well-defined intervention on exposure A. Suppose there is some underlying "version of treatment" variable K that takes values among some set $\mathcal{K}$, and that for each version of treatment k∈$\mathcal{K}$, the version is sufficiently well-defined to correspond to a unique potential outcome $Y_k$.[6,18] Suppose the investigator has access only to a coarsened variable A where each value of A corresponds to one or more values of K. We might then refer to the variable A as a "composite exposure" or "compound treatment" since each value of A can come about through numerous more specific "versions of treatment" K.[7,18,19] See Figure 1; the red arrows in this figure, and all subsequent figures, are those emanating from variables, related to the exposure, that are causally efficacious for the outcome.

We say there is no confounding for the effect of K on Y given C if $Y_k$ is independent of K conditional on C. We say that the consistency assumption holds if when K=k then Y= $Y_k$. Suppose the investigator has only information on (A,Y,C). Analogous to the formula above, it may seem natural to compute:

$$\sum_c \{E[Y|A = a, c] - E[Y|A = a^*, c]\} P(c).$$

However, it is not clear how to causally interpret this quantity when there are not well-defined interventions on A. Theory for multiple versions of treatment provides an interpretation.[6,7] It can be shown[6] that if the effect of K on outcome Y is unconfounded given C and if the consistency assumption holds, then

$$\sum_c \{E[Y|A=a,c] - E[Y|A=a^*,c]\}P(c)$$
$$= \sum_{k,c} E[Y_k|c]\, P(k|a,c)P(c) - \sum_{k,c} E[Y_k|c]\, P(k|a^*,c)P(c)$$

(1)

The first expression in equation 1 is the empirical quantity we would ordinarily use to estimate effects of A on Y, if multiple versions were not an issue. The second expression provides a causal interpretation. It can be interpreted as a comparison in a hypothetical randomized trial in which, within strata of covariates C, each individual in one arm is randomly assigned a "version of treatment" K from the underlying distribution of K in the subpopulation with (A=a,C=c), and each in the other arm is randomly assigned a "version of treatment" K from the underlying distribution of K in the subpopulation with (A=a*,C=c). An illustration with BMI and mortality is given in the eAppendix. While the original theory[6] assumed A constituted a coarsened version of K so that the relationship between K and A was a deterministic many-to-one mapping, this assumption can be weakened. As shown in the Appendix and discussed in the eAppendix, beyond unconfoundedness and consistency, the only assumption needed to derive the relation in 1 is that Y is independent of A conditional on (K,C) i.e. conditional on C, A gives no information about Y once K is known. The relationship between K and A then need not be many-to-one and can, moreover, also be stochastic. This will be important in the development that follows.

The MVT theory allows us to make formal progress in interpreting causal effects of composite exposures. There are, however, limitations to this approach. First, when the set of versions of treatment, $\mathcal{K}$, is unknown, this hinders precise understanding of the interpretation. Second, with the set of underlying versions unknown, it would then effectively be impossible to implement the hypothetical randomized trials embedded within the interpretation. Third, the interpretation will vary depending on what is included in C since, once C is fixed, this may limit the range of potential "versions of treatment" that are possible. Fourth, with the versions of treatment unknown, it becomes difficult to substantively assess the unconfoundedness assumption and thus to know whether the proposed interpretation is reasonable. Although the MVT interpretation has limitations, it may be the best we can do concerning a formal potential-outcomes-based interpretation of the quantitative effect estimate of a composite exposure,[18,19] and may also provide insights into where to focus intervention attempts (see eAppendix). In the next section, we will consider how this interpretation can be applied to measures arising from reflective or formative measurement models.[8-13]

**Causal Inference and Latent Variables: Reflective and Formative Models**

The classical model used in much measurement theory and scale development presupposes an underlying latent continuous variable η that gives rise to measured indicators $(X_1, \ldots, X_n)$ as in Figure 2a. After standardization, it is often assumed that each indicator $X_i$ is given by a linear function of η plus random error $\varepsilon_i$:

$$X_i = \lambda_i \eta + \varepsilon_i$$

(2)

The random errors $\varepsilon_i$ are often, but not always, assumed independent. This model forms the basis of much psychometric measure evaluation.[20] However, after this evaluation is complete, the measures that are used are generally just some function of the indicators $(X_1, \ldots, X_n)$. When the indicators are on the same scale, their mean is often used. Let $A = f(X_1, \ldots, X_n)$ denote the measure employed. This is typically considered an imprecise measure of the underlying latent $\eta$ that corresponds to the psychosocial construct of interest. Interest often then lies in assessing the relationship of this with various outcomes.

To estimate effects, often a regression is fit of Y on (A,C), assuming relationships depicted in Figure 2b:

$$E[Y|a,c] = \beta_0 + \beta_1 a + \beta_2' c$$

Provided the covariates C control for confounding, $\beta_1$ is then sometimes interpreted as the causal effect of the exposure on the outcome. Sometimes, especially when structural equation models are used, this estimation is done with correction for measurement error, using reliabilities $\lambda_i$ from the measurement model[4] and the estimate is then interpreted as the causal effect of the latent $\eta$ corresponding to the underlying construct. When the reliabilities $\lambda_i$ vary across indicators and this is neglected, as would often be the case if Y were simply regressed on (A,C), then this interpretation is problematic. However, even in this setting, a causal interpretation of $\beta_1$ is possible using MVT theory. Specifically, if we replace K in the previous section with $\eta$, and compare measure level A=a+1 to A=a, then if the effect of $\eta$ on Y is unconfounded given C, as in Figure 2b, equation 1 above becomes:

$$\beta_1 = \sum_{\eta,c} E[Y_\eta|c] P(\eta|A = a+1, c) P(c) - \sum_{\eta,c} E[Y_\eta|c] P(\eta|A = a, c) P(c).$$

In other words, even if reliabilities $\lambda_i$ vary and this is ignored, $\beta_1$ can still be interpreted as a comparison in a hypothetical randomized trial in which, within strata of covariates C, individuals in one arm are randomized to a value of $\eta$ from the actual distribution of $\eta$ in the subpopulation with (A=a+1,C=c), and individuals in the other arm are randomized to a value of $\eta$ from its actual distribution in the subpopulation with (A=a,C=c). We can apply the result from equation 1 with K replaced by $\eta$ because, in Figure 2b, $A = f(X_1, \ldots, X_n)$ will be independent of Y conditional on $(\eta,C)$.[15] Similar remarks concerning independence pertain to the other results below.

The model in Figure 2a is sometimes referred to as a reflective model because the indicators reflect the underlying latent variable. An alternative model for measurement is sometimes called the formative model[9-11] and is illustrated in Figure 3a. In this model, the indicators effectively together form the underlying variable of interest, which is a function of the indicators plus error:

$$\eta = \sum_i \lambda_i X_i + \varepsilon$$

(3)

In practice, measures are again formed as some function of the indicators $A = f(X_1, \ldots, X_n)$. Sometimes it is assumed that there is no error, and the function of the indicators is itself the underlying variable of interest with $\eta = A = f(X_1, \ldots, X_n)$. Considerations as to whether and when reflective or formative models are more appropriate are described elsewhere, though this continues to be debated.[9-13]

However, in this case also the causal interpretation under MVT theory is applicable. Provided the effect of η on Y is unconfounded given C, as in Figure 3b, then under the regression model $E[Y|a,c] = \beta_0 + \beta_1 a + \beta_2' c$ we again have:

$$\beta_1 = \sum_{\eta,c} E[Y_\eta|c] P(\eta|A = a+1, c) P(c) - \sum_{\eta,c} E[Y_\eta|c] P(\eta|A = a, c) P(c)$$

and $\beta_1$ can be interpreted as under the reflective model.

However, this analysis of reflective and formative models assumed that the latent η was causally efficacious. This may not be the case. Nothing in Figures 2a and 3a, nor in equations 2 and 3 require that it is η, rather than $(X_1, \ldots, X_n)$, that is causally efficacious. Consider instead the causal diagrams in Figures 4a and 4b. These correspond to reflective and formative models but with indicators $(X_1, \ldots, X_n)$, rather than the latent η, having causal effects on outcome Y. Importantly, the causal diagram in Figure 4a is compatible with the reflective model in Figure 2a and with equation 2. The causal diagram in Figure 4b is compatible with the formative model in Figure 3a and equation 3. We might thus distinguish between *basic* reflective and formative models represented in Figures 2a and 3a respectively (and by equations 2 and 3), versus what we might call *structural*[15,21] reflective and formative models, represented by Figures 2b and 3b respectively, which additionally assume that all causal relations with $(X_1, \ldots, X_n)$ are through η (reflective)[15,21] or that all effects of $(X_1, \ldots, X_n)$ are through η (formative). Both the structural models in Figures 2b and 3b, and also the models in Figures 4a and 4b, are compatible with the basic formative and reflective in Figures 2a and 3a.

In the causal models in Figures 4a and 4b, we might consider the effects of each indicator one-by-one. However, we might also instead consider measures formed as functions of the indicators $A = f(X_1, \ldots, X_n)$. If we regress Y on (A,C) using $E[Y|a,c] = \beta_0 + \beta_1 a + \beta_2' c$, we can again interpret the coefficient $\beta_1$ using MVT theory, this time taking K as the set of indicators $(X_1, \ldots, X_n)$. In both Figures 4a and 4b, the effects of $(X_1, \ldots, X_n)$ on Y are unconfounded conditional on C. For any two values A=a+1 and A=a, we can thus interpret $\beta_1$ by equation 1 with $K = (X_1, \ldots, X_n)$ and $A = f(X_1, \ldots, X_n)$. The coefficient $\beta_1$ can thus be interpreted as a comparison in a hypothetical randomized trial wherein, within strata of covariates C, individuals in one arm are randomly assigned to values of $(X_1, \ldots, X_n)$ from the actual distribution of these indicators in the subpopulation with (A=a+1,C=c), and individuals in the other arm are randomly assigned to values of $(X_1, \ldots, X_n)$ from the actual distribution of these indicators in the subpopulation with (A=a,C=c). The MVT interpretation is again applicable. However, now the interpretation extends to hypothetical interventions on the indicator set $(X_1, \ldots, X_n)$, rather than the underlying latent η.

We are left with the question of which of these causal models is more reasonable. Compatibility of the data with Figure 2a and equation 2, or with Figure 3a and equation 3, tells us nothing as to whether the indicators themselves, or some underlying latent variable, is causally efficacious. The next section presents analyses concerning associations between social integration and health suggesting that, in this case, a model with a causally efficacious univariate latent might not be plausible. Critically, *structural* formative and reflective models are incompatible with one of the indicators being causally related to the outcome and another not, because, under the structural models, that could only be the case if one of the $\lambda_i$ were 0, in which case it would not be an indicator of the underlying latent η at all.

**Social Integration Example**: Questioning the Measurement Models

Numerous studies have examined associations between measures of social integration and subsequent health. Social integration has been conceptualized and measured in a variety of ways.[22,23] However, evidence has been consistent across operationalizations that social participation tends to be associated with better health.[22]

Chang et al.[24] used data from the Nurses Health Study (n=76,362) to examine associations between social integration and incident coronary heart disease (CHD). They used a simplified Berkman-Syme Social Integration Index in 1992 as their exposure (summing indicators, each scored 0-3, of religious service attendance, community group participation, number of close friends, and marital status), and followed incident CHD through 2014, employing proportional hazards models. After adjusting for age, education, husband's education, census-tract income, hypertension, diabetes, cholesterol, family MI history, and depressive symptoms, comparing highest versus lowest quartiles of social integration, they estimate a hazard ratio of HR=0.79 (95%CI:0.70,0.88). Under the assumption of unconfoundedness, this association could potentially be interpreted under the MVT theory above. However, Chang et al.[24] also consider associations with each social integration indicator. They report evidence for an association between attending religious service more than once per week and lower CHD (HR=0.82; 95%CI:0.72,0.93), but no evidence for an association with other indicators, which all have point estimates close to HR=1. They report that associations are similar after adjusting for all indicators simultaneously. Similar conclusions were reached by Li et al.[25] and VanderWeele et al.[26] examining associations of social integration with all-cause mortality and suicide, respectively, with religious service attendance manifesting the strongest, or only, associations among the components (with marriage also protectively, but more weakly, associated with mortality).[25] Of course, these associations may still be confounded; moreover, the longitudinal associations of these indicators may differ with other outcomes such as happiness, income, prejudice, autonomy, etc. However, the present analyses suggest that we should be wary of assuming that "social integration" has a well-defined effect on a given outcome. From these prior analyses, it seems the indicators may be differentially associated with outcomes as in Figure 4. The *structural* formative model with a univariate latent is unlikely to hold.

**Further Critique of the Assumption of an Underlying Univariate Latent Variable in Reflective Models**

The assumption of a *univariate* causally relevant latent variable is strong and will often not correspond to reality. For reflective models, the assumption is sometimes defended on the grounds of factor analyses suggesting a unidimensional latent variable suffices to explain the covariance structure among indicators. But this does not entail a structural interpretation. The univariate factor model fitting a set of indicators (represented in Figure 2a and Equation 2) is consistent with a *structural* interpretation (Figure 2b) or with the latent being inert and the indicators being causally efficacious (Figure 4a). The goodness of fit of a unidimensional factor model in Figure 2a and Equation 2 tells us *nothing* about which of these causal models is a better representation of reality. Factor analysis may be useful in generating hypotheses about underlying causally efficacious univariate latent variables, but do nothing to establish this.

The structural interpretation of a reflective model is in fact empirically testable.[21] A structural interpretation would imply that randomized interventions that altered the latent η would have effects on the various indicators, $X_i$, that were proportionate to their reliabilities $\lambda_i$, which can be tested.[21] The structural reflective model is also incompatible with one indicator having an association with an outcome, and another not, or out of proportion with their reliabilities, which can also be tested.[21] The application of such tests to prominent scales, such as the Satisfaction with Life Scale (SWLS)[27] indicates that the structural interpretation can be rejected:[21] while there is evidence that 4 of 5 indicators are associated with all-cause mortality, for one indicator, "If I could live my life over, I would change almost nothing" there is effectively no such evidence.[21] We should be wary of assuming that a structural factor model always holds.

This does not imply the measure is bad, or that the basic univariate model is a bad fit for the covariance. The lack of a structural interpretation furthermore does not threaten the use of the scale as an outcome. Provided those using the scale equally value the individual indicators as outcomes, it is reasonable to take their average. Considered in this way, this average might then effectively be viewed as an index. Similar comments likewise pertain to other measures taken as outcomes. The interpretation of constructed measures as outcomes is arguably easier than as exposures. However, even as exposures, and even when the univaraite structural interpretation fails, the interpretation given by MVT theory is still applicable. But when used as an exposure, we must be careful, as even the MVT interpretation obscures the differential associations across indicators and obscures our capacity to discern the most relevant underlying constituents of reality. In this case, that the indicators themselves are differentially associated with all-cause mortality suggests there is no underlying *univariate* causally efficacious latent variable.

Similar remarks might well pertain to numerous other scales. The assumption of an underlying *univariate* structural factor is generally just presumed, not tested. Associations between different indicators and outcomes are rarely examined. The presumption of a *univariate* structural latent variable may well be unrealistic in numerous other settings.

**Towards More Complex Models**

From the analyses above, Figure 4 seems a better representation of reality than the structural reflective and formative models in Figures 2b and 3b. However, even Figure 4 is a gross simplification. Concerning social integration, each indicator for marital status, community group participation, number of close friends, and religious service attendance corresponds to a more complex reality. Quality of marriages vary; religious services can differ dramatically in content; community groups vary from arts to sports to card games. Thus, each indicator captures only aspects of a more complex reality, as in Figure 5, wherein the indicators $X_i$ each arise from potentially multidimensional underlying latents $\eta_i$. Models like this have been considered previously assuming univariate $\eta_i$.[12] But is it reasonable to assume each $\eta_i$ is univariate? Religious services vary in length of time, in discursive content, in style of worship, in demands made by participants, etc. Even the assumption that the "latent" behind a single indicator is univariate may be wrong. Nevertheless, the MVT interpretation of associations between measures A and outcome Y in terms of hypothetical randomized trials on $\eta = (\eta_1, \ldots, \eta_n)$ would still be applicable under Figure 5, and even so if each multivariate $\eta_i$ affected the entire set $(X_1, \ldots, X_n)$.

Additionally, the indicators themselves may vary over time; there will likely also be causal relations between the different aspects of the underlying reality (e.g. of social integration). Models representing this, either with only the indicators themselves, or with underlying latents also, are given in Figures 6 and 7, respectively. In these figures, several things become apparent. First, if control is not made for all indicators simultaneously then the associations of one indicator may confound that of another. For example, in Figure 6, suppose there were no effect of $X_1^t$ on Y; if we do not control for $X_n^t$ in a regression of Y on $X_1^t$, we might observe an association between the two simply because $X_n^{t-1}$ affects $X_1^t$, and $X_n^{t-1}$ also affects $X_n^t$ which affects Y. Second, the use of an indicator at a single time-point may be capturing, however crudely, the associations of an entire history of social participation. If we use a single composite measure that is a function of the indicators at a single time-point $A = f(X_1^t, \ldots, X_n^t)$, then if temporally prior levels of these indicators $(X_1^{t-1}, \ldots, X_n^{t-1})$ are causally related to the outcome Y independent of $(X_1^t, \ldots, X_n^t)$, either directly (Figure 6) or through the latents (Figure 7), then the associations between A and Y may also partially reflect associations of the outcome with past indicators $(X_1^{t-1}, \ldots, X_n^{t-1})$. Third, considerations of confounding control must take into account the time-varying nature of the indicators (and/or latents).[28] The MVT interpretation requires control of confounding for the underlying "versions-of-treatment" variable K. If K corresponds to a historical trajectory, its time-varying nature must be accounted for in confounding control. Unfortunately, confounding considerations become more complex with time-varying exposures[28,29] and if confounders can themselves be affected by prior exposure levels, traditional regression-based adjustment for confounding fails; more sophisticated models are needed.[28,29] A possibly attractive alternative is using the indicators at time t in the analysis, while simultaneously controlling for past values of the indicators (along with confounders C) at time t-1.[29,30] This proposal is discussed further in the eAppendix.

**Towards a New Model of Measure Construction: Constructs, Indicators, Measures, and the Underlying Constituents of Reality**

Abstracting yet further from the diagrams above (and setting aside the conditioning variables), there is arguably some complex underlying reality ($\mathscr{R}$ in Figure 8). Certain aspects of this constitute exposure states η related to the construct of interest. The multi-dimensional variable η takes values in some set $\mathscr{K}$, each member of which defines a potential outcome for outcome Y. The variable η corresponds to the "version-of-treatment" variable K in the MVT theory; η is multivariate, not univariate. This multi-dimensional η gives rise to a set of observed indicators $(X_1, \ldots, X_n)$, from which we form measures $A = f(X_1, \ldots, X_n)$, either as a mean, or some other function arising from measure development processes and psychometric evaluation. We use either the indicators $(X_1, \ldots, X_n)$ or the summary measure A in analyses and examine associations with outcomes of interest, Y, controlling for other covariates, and possibly past values of $(X_1, \ldots, X_n)$, or A, as appropriate.

Concurrent with these processes giving rise to indicators and measures, is the process by which we form our concepts and constructs. The underlying constituents of reality, and our living as persons within communities, gives rise to our language and the concepts embedded within it. In order to try to systematize and study various aspects of the underlying reality, we propose constructs. Such constructs characteristically involve the systemization and reduction of our ideas, language, and concepts so as to operationalize them for use in specific modes of reasoning. However, language itself, and the concepts and derivative constructs embedded in it, of course go on to shape human behavior, the items and measures we propose, and study participants' responses to them. These two processes are represented diagrammatically in Figure 8. In constructing measures, we hope that our measures correspond to our constructs.

The dominant measurement models – the reflective models in Figure 2 and formative models in Figure 3 – each capture aspects of these processes, but each arguably fails to acknowledge important features. Formative approaches get right that our measures are always functions of our indicators. Our measure of social integration is formed by our indicators; it is not that there is a true univariate "social integration" that itself causes the indicators. However, formative models misconstrue the relationship between our measures and the underlying reality to be studied. It is not that our measures, formed by the indicators, constitute (possibly subject to error) the underlying reality to be studied (as in Figure 3). It is the underlying reality that gives rise to our indicators by which we form measures.

Reflective models, in contrast, get right the fact that our measured indicators do not cause the relevant constituents of reality under study, but rather are caused by, or reflective of, these features. However, reflective models are wrong in equating the relevant aspects of reality with a univariate latent variable that corresponds to our construct.[31-33] There is no underlying *univariate* latent variable that corresponds to our construct, say, of intelligence, such that "true intelligence" gives rise to the measured indicators. The underlying reality corresponding to our constructs is far more complex than a univariate variable. Models that use multiple latent variables[12,Figure-5] more closely correspond to the underlying processes but still wrongly equate reality to a few univariate latents.

Thus, even in paradigmatic cases of the formative model, such as social integration, concerning which there is no true underlying social integration variable that "causes" the indicators, the underlying reality is nevertheless more complex than the social integration measure. Likewise, even in paradigmatic cases of the reflective model such as intelligence, concerning which the indicators of test responses do not "cause" intelligence, it is still the case that the underlying reality is again more complex than a univariate general intelligence latent variable.[33,34] In both cases, a complex underlying reality gives rise to our indicators from which we form measures. This is true even if we develop those measures based on psychometric approaches, such as factor analysis, arising from reflective models. Even then, the measures are still ultimately functions of the indicators, as in Figure 8. If we lose sight of that fact, we may forget that certain indicators, corresponding to particular aspects of the underlying reality, may in fact be differentially related to our outcomes of interest (as in Figures 4-7).

These issues likewise pertain to distinctions drawn between "scales" with closely related items (supposedly corresponding to reflective models) and "indices" with items that are conceptually distinct but somehow together form the construct of interest (often thought to correspond to formative models). The model for measure construction given in Figure 8 is arguably applicable to both scales and indices. In both cases, a complex underlying reality gives rise to item responses from which we form measures. The relations between the underlying processes and the formation of measures may thus be more similar for scales and indices than typically thought. Whether a measure is considered a scale or index may have more to do with the items used to construct the measure, and the use of that measure, than with the definition of the construct itself. While life satisfaction is often assessed as a "scale" with several related subjective indicators[27], if the construct is instead assessed by life domain (work, family, health, finances, etc.)[35,36] the measure will resemble an index. Conversely, while measures of social integration are often assessed by domain (marital status, time with friends, religious community, etc.)[22,23], social integration could alternatively be assessed with a series of related subjective indicators as in Duke's Subjective Social Support subscale.[37] In all these cases, a complex underlying reality gives rise to indicators by which we form measures. It is the conceptual relations between the items and the construct that differs, not the model of measurement per se.

All measures are formative in that they are formed from observed indicators; all measures are reflective in that they are reflective of a more complex underlying reality. The fallacy of the formative model is that the relevant underlying reality is made up of a function of our indicators; the fallacy of the reflective model is the supposition that we have imperfectly measured an underlying *univariate* latent variable.

**Implications and Conclusions**

The dominance of the reflective model, and the fallacious presumption that the basic univariate factor model fitting the indicators well implies a structural interpretation, gives rise to the illusion that, in most settings, there truly is an underlying *univariate* latent variable adequately representing reality.[21,38] There is no reason to assume this is true. But this presumption and illusion has arguably led to a related series of other subtle subsequent missteps in measure construction, conceptualization, and evaluation.

It has been argued elsewhere that current factor analysis and measure construction practices have led to the conflation of the terms "construct" and "latent variable."[32,39] Indeed the very term "latent construct" effectively entails the equating of a conceptual specification with a quantitative variable, generally presuming a univariate structure. We need a clear distinction between concepts and constructs, their underlying referents ($\eta$), and our attempts to measure these underlying referents ($X_1, \ldots, X_n$).[32,39] The conflation of "construct" and "variable," and the presumption of a univariate underlying reality has also led to a notion that the nature of the concept is to be discovered empirically from analyses of correlations.[32] Items are proposed, factor analyses implemented, and it is assumed we somehow thereby come to understand the meaning of the construct itself. This view has in turn has often led to a lack of formal definitions given for the construct under consideration[32], since, so it is thought, this is to be "discovered" empirically. While plenty of theory is often provided, formal definitions are rare.

The lack of definitions in turn obscures the relationship between items and constructs. Items that are necessary or sufficient or merely illustrative of the construct are treated interchangeably; none are related to definitions themselves. Without definitions, it becomes difficult to assess whether two different measures of allegedly the same construct are intended to assess the same thing, or whether authors have different understandings of the construct, or whether they view the nature of the construct as something to be discovered empirically and are beginning exploration from different places. The lack of definitions also tends to lead to overly broad inclusion of items within measures. Not infrequently, conceptually distant but desirable outcomes are placed among the items (e.g. "I've been pretty successful in life" in Synder's *hope* scale[40]; or taking "Pride in your achievements" in the Connor-Davidson *resilience* scale[41]). Without definitions, criticism is more difficult; often these items are simply accepted, provided a univariate factor model accounts for covariances among indicators. Moreover, since the underlying factor is presumed univariate, item-by-item analyses, which might uncover differential relationships with outcomes, are rare, thereby further obscuring the important conceptual and empirical distinctions that may be present among the items.

Much of this would benefit from change, beginning with clear definitions of the construct.[42] Proposed items should then be derived from the definitions, with an understanding of their relationship including whether items make use of the word corresponding to the construct; whether the items are necessary, sufficient, necessary and sufficient, or merely illustrative for the construct; or whether items are intended to capture different facets of the construct. The work of analytic philosophy may be useful both in this task, and in clarifying different uses of our language and thereby facilitating particular definitions of the construct in view.[43-45] Various measures can be proposed from item indicators on conceptual grounds. Appropriate cognitive testing and measure evaluation strategies could be developed. Factor analyses[1-4] may be useful to assess approximate covariance dimensionality, but indication of unidimensional factor structure is neither necessary nor sufficient for using a univariate measure in analysis. It is not necessary because associations between constructed measures and outcomes can still admit a causal interpretation under the MVT theory[6] above. It is not sufficient because even if the basic univariate factor model fits, the causal interpretation of the latent may not be structural.[21] Regardless of the fit of the unidimensional model, it will still be useful to carry out item-by-item analyses, or using composites of conceptually-related items, either to potentially provide some evidence for a *structural* univariate latent interpretation, or alternatively to uncover important distinctions between items that may be relevant in refining measure construction, understanding facets of the construct, or in thinking about interventions.

A preliminary outline of a more adequate approach to the construction and use of psychosocial measures might thus be summarized by the following propositions, that I have argued for in this paper: (1) Traditional *univariate* reflective and formative models do not adequately capture the relations between the underlying causally relevant phenomena and our indicators and measures. (2) The causally relevant constituents of reality related to our constructs are almost always *multidimensional*, and from these we assess indicators and construct measures, and from the underlying reality our language and concepts also emerge, from which we propose more precisely defined constructs. (3) In measure construction, we ought always specify a *definition* of the underlying construct, from which items are derived, and by which analytic relations of the items to the definition are made clear. (4) If a structural interpretation of a univariate reflective factor model is being proposed this should be *formally tested*, not presumed; factor analysis is not sufficient for assessing the relevant evidence. (5) The *presumption* of a structural univariate reflective impairs measure construction, evaluation, and use. (6) Even when the causally relevant constituents of reality are multidimensional, and a univariate measure is used, we can still interpret associations with outcomes using theory for *multiple versions of treatment*, though the interpretation is obscured when we do not have a clear sense of what the causally relevant constituents are. (7) When data permit, examining associations *item-by-item*, or with conceptually related item sets, may give insight into the various facets of the construct.

A new model of measurement for psychosocial constructs is needed in light of these points – one that better respects the relationships between our constructs, items, indicators, measures, and the underlying causally relevant phenomena.

**Appendix**

*Causal Effects Under Multiple Versions of Treatment*

The derivation here follows the structure of the proof of Proposition 8 given by VanderWeele and Hernán[6] but under weaker assumptions (see eAppendix), requiring only that (i) Y is independent of A conditional on (K,C); (ii) the effect of K on Y is unconfounded given C i.e. $Y_k$ is independent of K given C; and (iii) the consistency assumption that when K=k then $Y_k$=Y. Under these assumptions we then have that:

$$\sum_c \{E[Y|A=a,c] - E[Y|A=a^*,c]\}P(c)$$
$$= \sum_{k,c} E[Y|k,a,c]\,P(k|a,c)P(c) - \sum_{k,c} E[Y|k,a^*,c]\,P(k|a^*,c)P(c)$$
$$= \sum_{k,c} E[Y|k,c]\,P(k|a,c)P(c) - \sum_{k,c} E[Y|k,c]\,P(k|a^*,c)P(c)$$
$$= \sum_{k,c} E[Y_k|k,c]\,P(k|a,c)P(c) - \sum_{k,c} E[Y_k|k,c]\,P(k|a^*,c)P(c)$$
$$= \sum_{k,c} E[Y_k|c]\,P(k|a,c)P(c) - \sum_{k,c} E[Y_k|c]\,P(k|a^*,c)P(c)$$

where the first equality follows from the law of iterated expectations, the second from the independence of Y and A condition on (K,C), the third from consistency, and the fourth from unconfoundedness.

**Figure Captions**

Figure 1. A model for multiple versions of treatment wherein the version-of-treatment variable K affects the outcome Y but is confounded by measured covariates C, with the measured exposure variable A representing a coarsening of K. [Red arrows in this figure and in all subsequent figures are those emanating from variables related to the exposure that are causally efficacious for the outcome.]

Figure 2a. Basic reflective model with univariate latent variable η giving rise to indicators $(X_1, \ldots, X_n)$.
Figure 2b. Structural reflective model with measure A as a function of the indicators $A = f(X_1, \ldots, X_n)$ and with all causal relations concerning the indicators from prior variables C, or outcomes variables Y operating through latent variable η.

Figure 3a. Basic formative model with the indicators $(X_1, \ldots, X_n)$ giving rise to a univariate latent variable η.
Figure 3b. Structural formative model with all causal relations with subsequent outcomes variables Y operating through latent variable η.

Figure 4a. Basic reflective model but with the causal relations from prior variables C or outcomes Y operating directing through the indicators $(X_1, \ldots, X_n)$ rather than the latent η.
Figure 4b. Basic formative model but with causal effects of the indicators $(X_1, \ldots, X_n)$ on outcome Y not through the latent η.

Figure 5. Multidimensional latent model with each indicator $X_i$ used in forming measure A arising from a potentially multidimensional latent variable $\eta_i$ which is causally efficacious for outcome Y. [Measured covariates C have been omitted for diagrammatic simplicity]

Figure 6. A model depicting the indicators, $X_i^t$ used to form measures A, themselves changing over time and causally affecting one another and the outcome Y [Measured covariates C have been omitted for diagrammatic simplicity]

Figure 7. A model depicting potentially multivariate latents $\eta_i^t$ giving rise to indicators $X_i^t$ from which the measure A is formed, with the latents themselves changing over time and affecting one another as well as the outcome Y [Measured covariates C have been omitted for diagrammatic simplicity]

Figure 8. A proposed new model of measure construction wherein complex underlying reality, $\mathscr{R}$ contains certain aspects of this reality (represented by the multi-dimensional variable η) relevant to the construct. These relevant aspects of reality give rise to a set of observed indicators $(X_1, \ldots, X_n)$, from which we form a measure A. [The dotted arrows, while in some sense causal, correspond to those relationships that are not explicitly between *variables*]

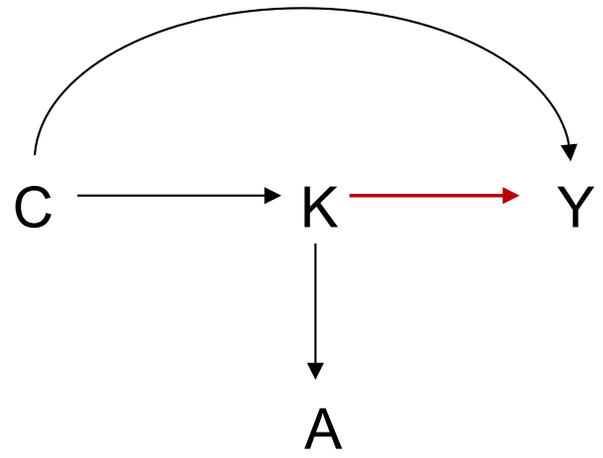

Figure 1

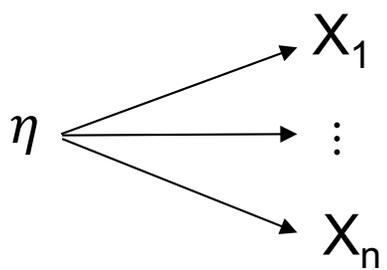
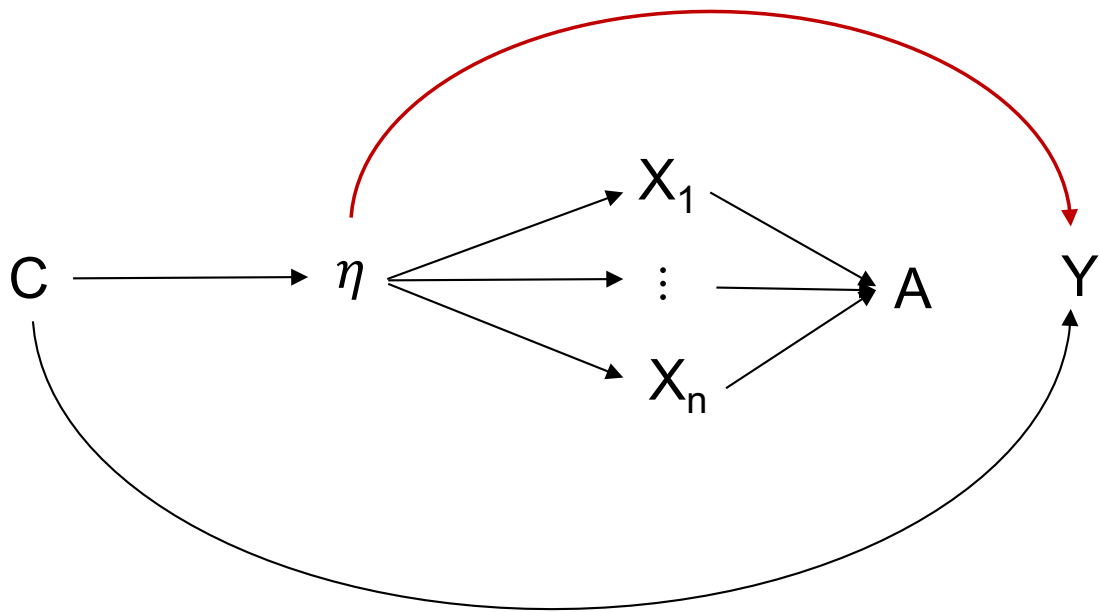

Figure 2(a)

Figure 2(b)

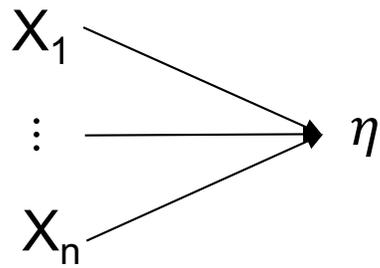

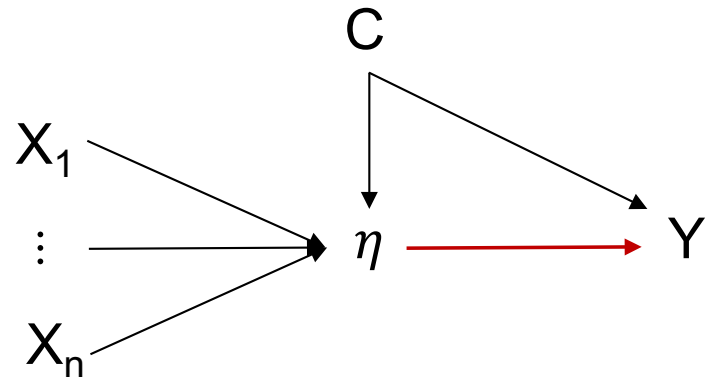

Figure 3(a)

Figure 3(b)

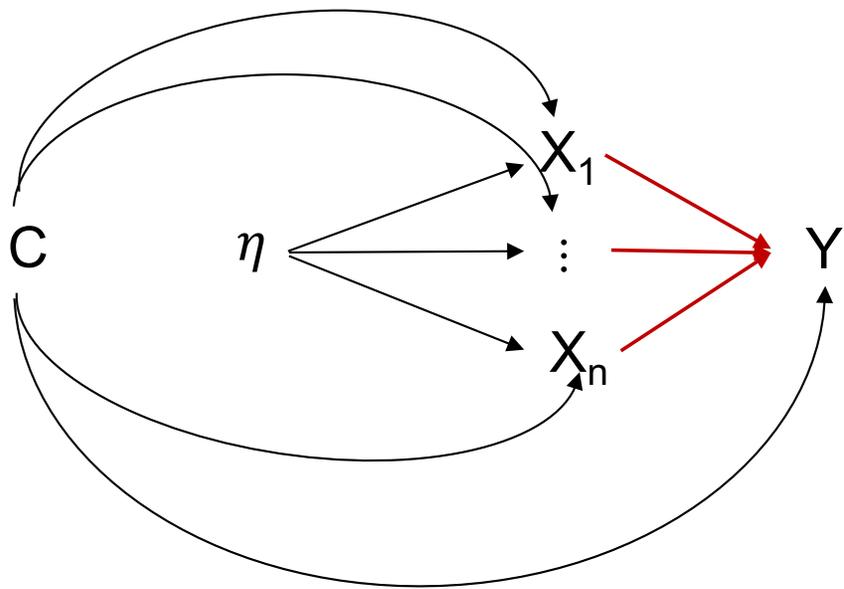 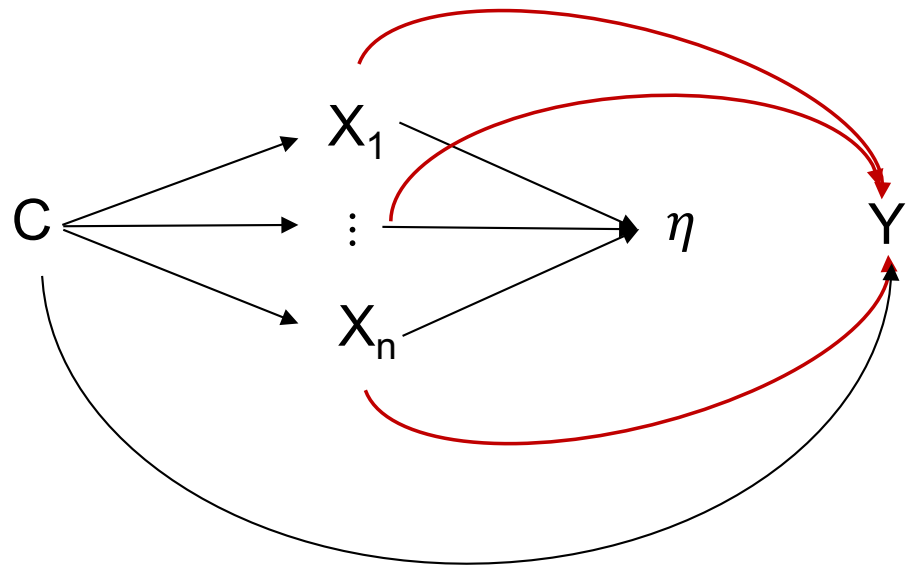

Figure 4(a)            Figure 4(b)

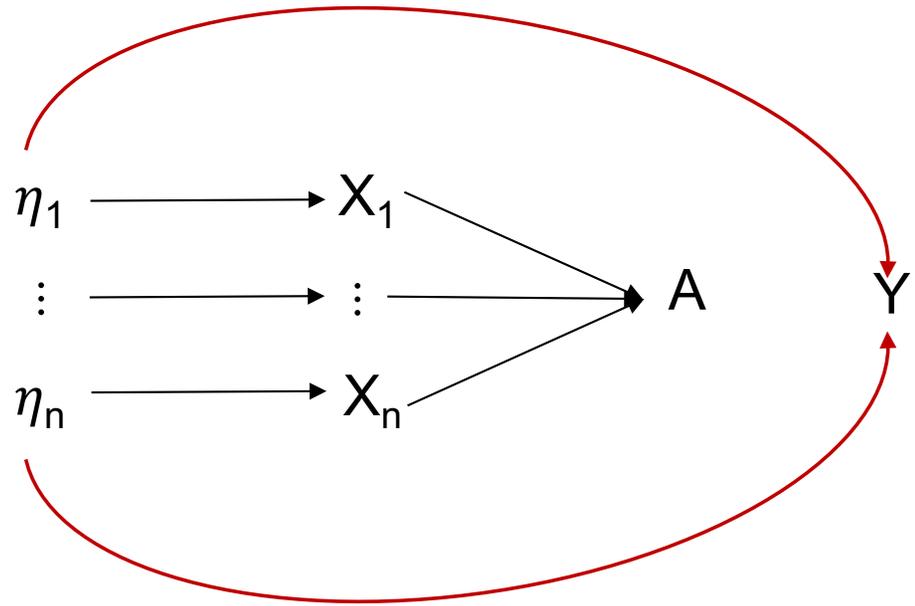

Figure 5

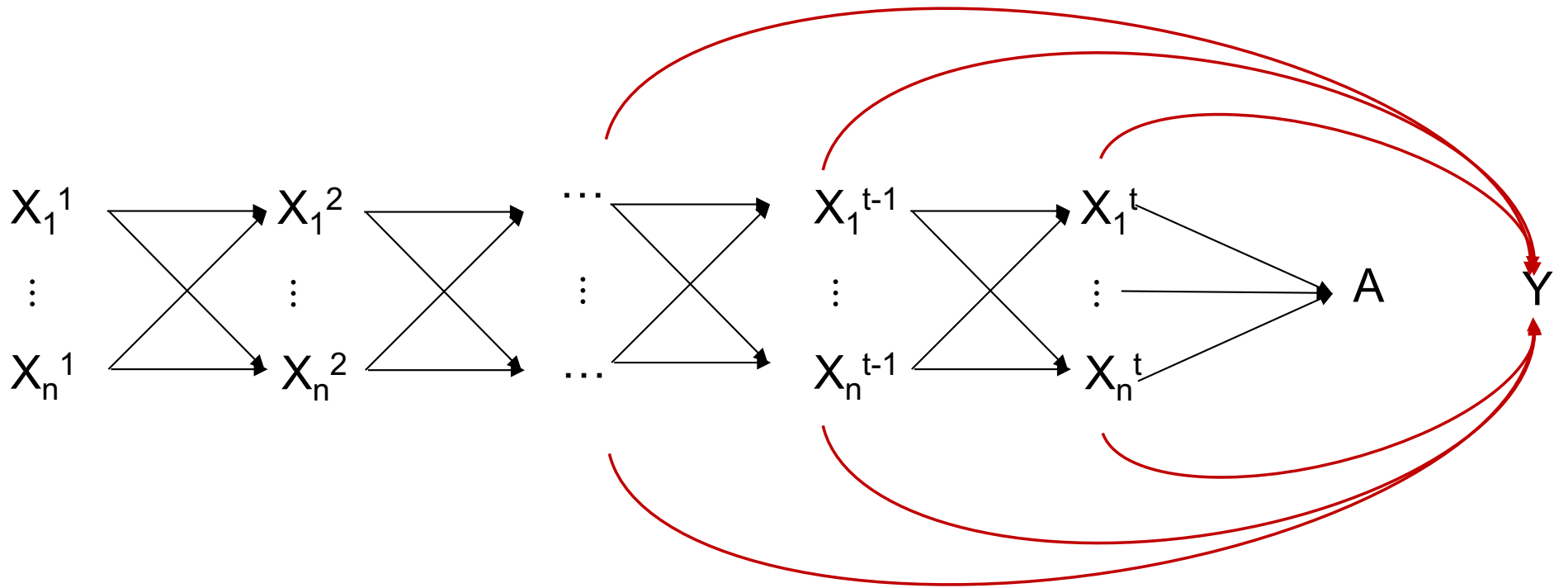

Figure 6

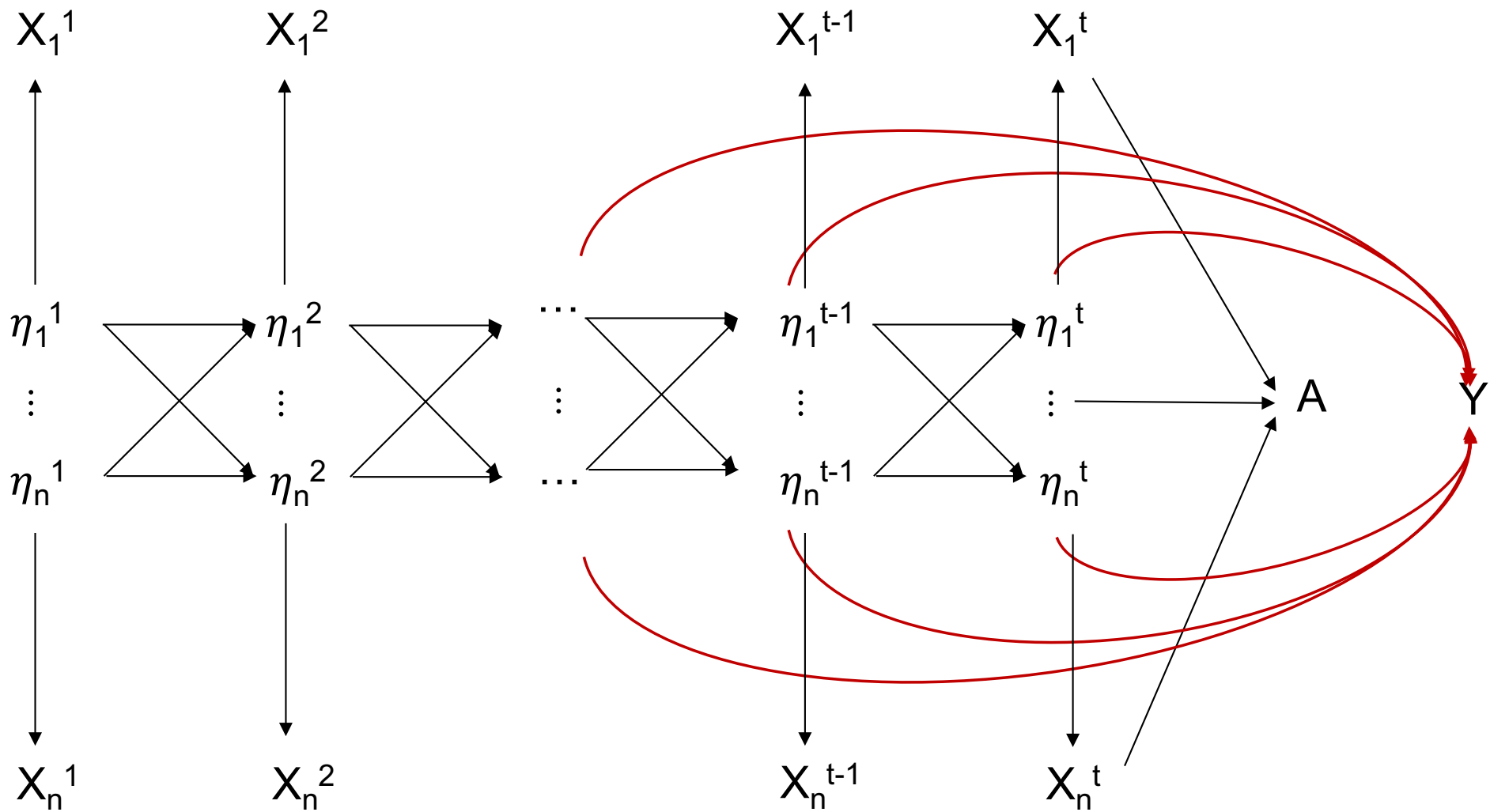

Figure 7

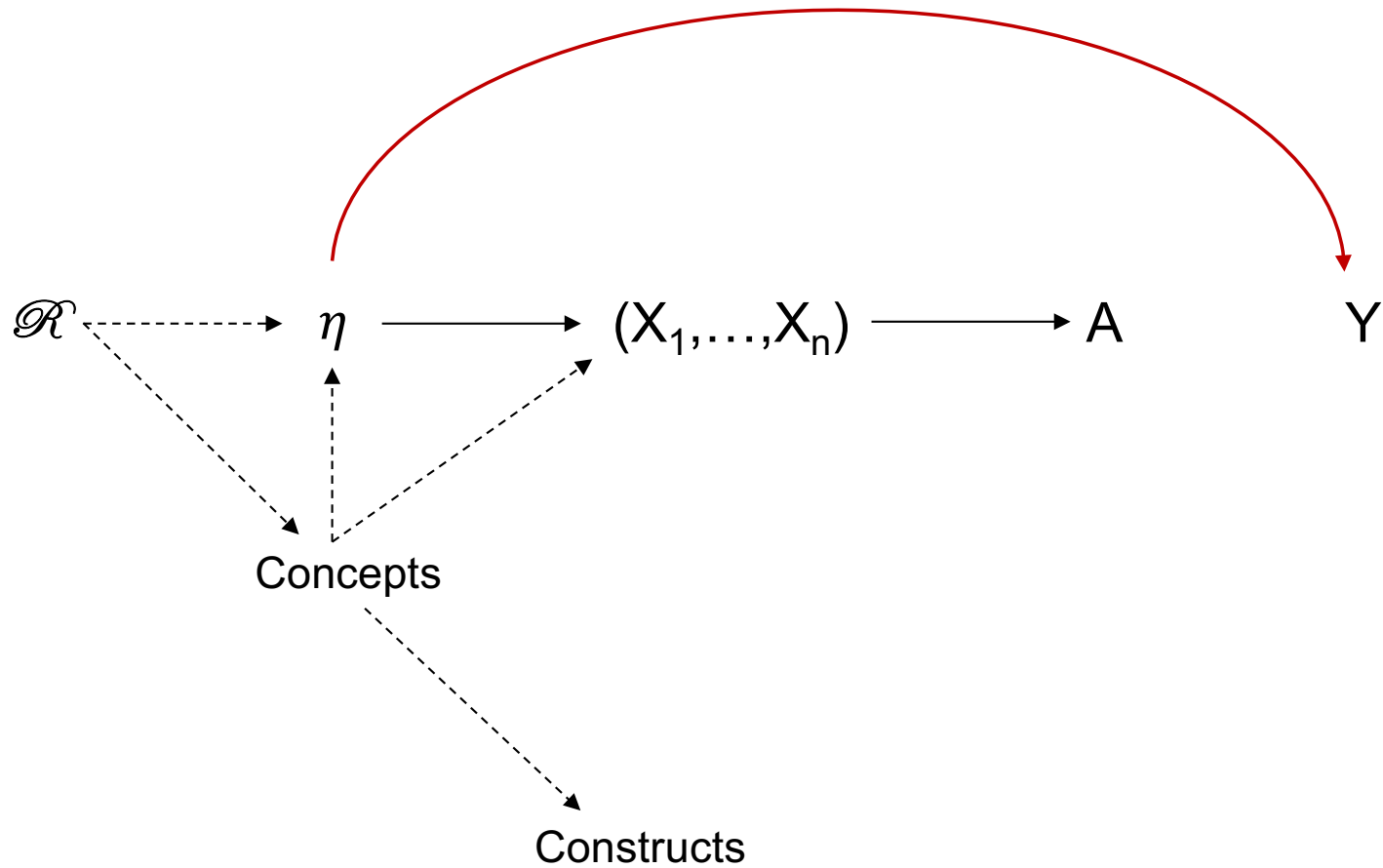

Figure 8

# eAppendix for "Constructed measures and causal inference: towards a new model of measurement for psychosocial constructs"

*Example of Causal Inference Under Multiple Versions of Treatment: BMI and Mortality*

One paradigmatic example to which multiple version of treatment (MVT) theory is applicable concerns attempts to assess the effect of being overweight or obese, as assessed by body mass index (BMI), on for example 10-year mortality risk.[6,7] There is no unique hypothetical intervention on BMI. One might accomplish a change in BMI from 30 to 25 by exercise, or dietary changes, or surgery, or through other means. Each of these might have very different effects on mortality risk. It then becomes difficult to speak unambiguously of *the* effect of BMI on mortality. The theory of causal inference under multiple versions of treatment was intended to provide a more precise interpretation of quantitative causal effect estimates in settings like this.[6,7]

Consider the setting in which A=1 corresponds to BMI=30 and A=0 corresponds to BMI=25 and Y is mortality after 10 years. In the case of BMI, a particular "version of treatment" k might correspond to a particular set of life-style choices from study entry onwards, which would in turn lead to a particular level of BMI.

To illustrate and make more concrete this interpretation, consider data analyses presented by Bhaskaran et al. (2018). In their paper, they used UK primary care data on 3.6 million persons from the Clinical Practice Research Datalink (CPRD), linked to national mortality registration data from January 1998 through March 2016. For never-smokers aged 16 and above, they used proportional hazards models for mortality, modeling BMI with splines and additionally controlling for baseline age, sex, alcohol use, diabetes, socioeconomic status, and calendar period. Their estimated hazard ratio comparing BMI=30 to BMI=25 was HR = 1.21 (95% CI: 1.20, 1.22). As noted above, there is no well-defined intervention on BMI. The underlying "versions of treatment" K for each person may correspond here to a complex trajectory of lifestyle choices that brings an individual to a particular BMI upon study entry. Full information on such lifestyle trajectories is of course unavailable in the data and it is effectively impossible to even fully enumerate all the possible trajectories. However, by the MVT result in the text, if we were willing to assume that the covariates in the model controlled for confounding of the effect of version of treatment on the outcome, then we could still potentially interpret the HR=1.21 using the MVT theory. The estimate HR of 1.21 would then correspond to the hazard ratio we would obtain in a hypothetical randomized trial which, within strata of covariates C, the participants in one arm are further each randomly assigned a lifestyle trajectory "version of treatment" K from the actual distribution of K in the subpopulation with BMI=30 and C=c, and in the other arm, persons are instead further randomly assigned a lifestyle trajectory "version of treatment" K from the actual distribution of K in the subpopulation with BMI=25 and C=c.

*Multiple Versions of Treatment Theory and Nested Randomization*

In many analyses of BMI, it is categories of BMI (e.g. 20-25 versus 25-30) that are compared, rather than actual values of BMI (e.g. BMI=30 versus BMI=25). The MVT theory can also be applied to categories of the exposure variable using a nested randomization. For example, if A=1 corresponded to BMI in the range of 25-30 and A=0 corresponded to BMI in the range of 20-25, then the estimate under the MVT theory could be interpreted as follows. Assuming the covariates controlled for confounding, the estimate would correspond to the hazard ratio we would obtain in a hypothetical randomized trial which, within strata of covariates C, the participants in one arm were first each randomly assigned to a BMI value between 25 and 30 according to the distribution of BMI between 25 and 30 from the actual subpopulation with BMI in the range 25-30 and covariates C=c and then further randomly assigned a lifestyle trajectory "version of treatment" K from the actual distribution of K in the subpopulation with C=c and the level of BMI to which they had been assigned; and in the other arm, persons would instead be randomly assigned to a BMI value between 20 and 25 according to the actual distribution of BMI between 20 and 25 from the subpopulation with BMI in the range 20-25 and covariates C=c and then further randomly assigned a lifestyle trajectory "version of treatment" K from the actual distribution of K in the subpopulation with C=c and the level of BMI to which they had been assigned. The MVT interpretation would still potentially be applicable but becomes yet more complex.

Likewise, in the social integration example given in the paper, the MVT interpretation would require nested randomization wherein, within strata of covariates C=c, for a given social integration quantile, the participants would be randomized to a value of the components of social integration (religious service attendance, community group participation, number of close friends, and marital status) according to the actual distribution of the components in that quantile among those with C=c, but then would be further randomized to a set of life choices and relationship trajectories K (including e.g. quality of relationship, type of community, style of religious service etc., represented by multi-dimensional $\eta$ in Figures 5 and 7) drawn from the actual distribution of these "versions of treatment" K among the subpopulation with actual levels of religious service attendance, community group participation, number of close friends, and marital status that they were hypothetically randomized to and with covariates C=c.

*Limitations of the Multiple Versions of Treatment Causal Interpretation*

As discussed in the text, the MVT interpretation is subject to a number of limitations and challenges. First, when the set of versions of treatment, $\mathcal{K}$, is unknown, this limits the precise understanding of the interpretation. One does not know all the various complex trajectories that may give rise a particular value of the measure A, nor, of course, does one then know the distribution of these various trajectories. Second, and relatedly, with the set of underlying versions unknown, it would then effectively be impossible to implement the hypothetical randomized trials embedded within the interpretation. The usefulness of the interpretation is thus, in this regard, somewhat limited. However, as discussed further below, the causal MVT interpretation may nevertheless provide clues as to where to best intervene. Third, the interpretation will vary depending on what is included in the measured covariates C. This is because, once C is fixed, this may limit the range of potential "versions of treatment" that are possible. For example, suppose in the context of BMI, exercise was included in the measured covariates C that were adjusted for in the analysis. In this case, because the hypothetical randomized trial in the interpretation is stratified by the measured covariates C, the distribution of the lifestyle trajectory "version of treatment" variable K will no longer vary by exercise across the BMI arms (or by whatever aspect of it is captured by the measured covariate) because the hypothetical randomized trial is effectively stratified by exercise. Fourth, with the versions of treatment unknown, it becomes difficult to substantively assess the unconfoundedness assumption and thus to know whether the proposed interpretation is reasonable.

Although the MVT interpretation has limitations, it may be the best we can do with respect to a formal potential-outcomes based interpretation of the quantitative effect estimate of a composite exposure.[18,19] Even reflective and formative models that assume an underlying univariate latent variable face the challenge of what one means by intervening upon the latent variable, and different interventions on it may again result in different causal effects on the outcome. Moreover, these reflective and formative models typically also additionally impose what is the often unrealistic assumption of an underlying *univariate* latent variable, which the MVT approach circumvents. Another alternative is of course abandoning a causal interpretation entirely, but what that often results in, in practice, is that the researcher/author/analyst/reader implicitly imposes a vague ill-defined causal interpretation. The MVT interpretation at least makes clear – and forces the interpreter to think about – the caveats.

While the limitations of MVT interpretation are important and need to be taken seriously, attempting to interpret associations causally, even if imprecisely, can still sometimes be valuable with regard to gaining insight into potential interventions. Analyses that suggest causal effects of phenomena related to our psychosocial constructs may help identify potential intervention targets. Attempts at estimating causal effects of phenomena related to our constructs of interest (e.g. using MVT theory) can be useful in informing on what to try to intervene first, i.e. concerning what we ought to begin with as potential intervention targets, whilst keeping us mindful (through the very limitations and assumptions of the MVT interpretation) that there are indeed likely multiple complex trajectories involved, and that the effects of our actual interventions may thus not correspond to what we had estimated. Nevertheless, when attempting to develop interventions to improve population health and well-being, one must begin somewhere. Attempting, as best as we can, to estimate causal effects of the phenomena of interest may provide important clues as to where to begin. For example, analyses that suggest that measures of purpose in life have considerably stronger associations with all-cause mortality than do measures of affective happiness (Trudel-Fitzgerald et al., 2019) might suggest that we ought to focus on the former as a potential intervention target. Of course, decisions concerning intervention development are shaped not only by the potential causal effects of the intervention target, but also by the plausibility that that target can be changed, and by the costs and ethical considerations of trying to change it. But analyses examining putative causal effects, as best as possible, that correspond to phenomena related to our constructs of interest may be useful information that can help inform such decision-making.

*Analysis and Interpretation of Associations with Time-Varying Indicators or Latents*

As noted in the text, in the context of time-varying indicators or latents, a possibly attractive analysis alternative is using the indicators at time t as the exposure, while simultaneously controlling for past values of the indicators (along with confounders C) at time t-1.[29,30] At least in Figure 6 this has the advantage of simplifying confounding control to avoid the potential complications arising from time-dependent confounding. In Figure 6, each of the indicators at time t could be considered one at a time as the exposure, while simultaneously controlling, when data allow, for past values of the entire set of indicators ($X_1^{t-1}, ..., X_n^{t-1}$). The associations between the time t indicators are then more easily interpreted as corresponding to the effects of present exposure on the outcomes. Moreover, control for prior exposure levels can help rule out reverse causation by prior outcome[29,30] and potentially helps rule out residual unmeasured confounding, since an unmeasured confounder would have to substantially affect current exposure through pathways independent of prior exposure to generate considerable bias. These remarks pertain directly to Figure 6 with causally efficacious indicators, or approximately in Figure 7, to the extent that the indictors capture what is causally relevant in the underlying reality with respect to the outcome Y. Unfortunately, relatively large sample sizes will be needed to deal with the potential collinearity arising between current and prior exposure levels, especially if the individual indicators themselves are being used. In cases of tens of thousands of participants, as in the Nurses' Health Study described in the text, this may be possible. The analyses of Li et al.[25] and VanderWeele et al.[26] discussed in the text in fact did control for past exposure levels. In principle, in Figure 6, it would be possible to regress the outcome Y on the entire set of indicators at time t, $(X_1^t, \ldots, X_n^t)$, while also controlling for the indicators at time t-1, $(X_1^{t-1}, \ldots, X_n^{t-1})$, but this would require even larger sample sizes to deal with collinearity. Another alternative, when sample sizes are more limited, would be to control for a summary measure of the indicators, e.g. their mean, at time t-1, $f(X_1^{t-1}, \ldots, X_n^{t-1})$. This will only partially control for confounding by the past indicators, but, depending on how strongly correlated the indicators are, it may suffice to remove most of the confounding. Which of these strategies is to be preferred will depend on a combination of the sample size available, the correlation among the indicators, and the extent to which there are differential effects of the indicators on one another and on the outcome of interest.

In summary, to obtain the cleanest possible interpretation of causal effect estimates, it may be desirable to regress outcome Y on the individual indicators one-by-one, along with past values of these indicators, and potentially confounding covariates. Associations with summary measures might still be interpreted as under the MVT theory but, as seen in the social integration example, this has the potential to obscure the more subtle relationships concerning each indicator.

*Discussion of the Independence Assumption for Multiple Versions of Treatment*

The independence assumption in the text and in the derivation given in the paper's print Appendix was that Y is independent of A conditional on (K,C). This was a weakening of the assumption employed in Proposition 8 of VanderWeele and Hernán[6] on the original multiple versions of treatment theory which was that the mapping from K to A was a deterministic many-to-one map. The assumption is weaker insofar as if the mapping from K to A was a deterministic many-to-one map then it immediately follows that Y is independent of A conditional on (K,C) since, conditional on K, A is a constant. However, the weakening of this assumption to Y being independent of A conditional on (K,C) also gives rise to ways that this assumption can be violated that would be precluded if the mapping from K to A were a deterministic many-to-one map. For example, if in Figure 2(b), with K=η, there were causal effects of one or more of the indictors $(X_1, \ldots, X_n)$ on Y, then the assumption would be violated. This might arise, for example, if the actual measurement of a particular indicator (or e.g. a physician's comment upon it to a patient) could potentially lead to change in behavior that affected the outcome itself. Likewise, the assumption would be violated if there were a common cause U that might affect one or more of the indicators $(X_1, \ldots, X_n)$ and also the outcome Y. Such unmeasured common causes might include, for example, intelligence, education, or ethnicity. These possibilities, however, do not ultimately threaten the applicability of the MVT interpretation insofar as K can itself be defined to include one or more of the indicators $(X_1, \ldots, X_n)$ or the common causes U of the indicators $(X_1, \ldots, X_n)$ and the outcome. The underlying constituents of reality η that are relevant for the outcome can all be defined as being part of the multivariate latent variable K in the MVT theory. By defining K in this manner, one effectively guarantees the validity of a causal diagram in which there no causal effects from the indicators $(X_1, \ldots, X_n)$ on the diagram to the outcome Y, since, if these effects were there, they would already be captured by the arrow corresponding to the causal effect from K=η to Y. The mediated or indirect effect of K=η on Y through $(X_1, \ldots, X_n)$ is effectively 0 since these effects are captured by the direct effect of K=η on Y. This is so by the very definition of K=η. Note that in the application of the MVT theory to the models in Figure 4 which had causal effects of the indicators themselves on the outcome Y, the variable K was taken to be the entire set of indicators $(X_1, \ldots, X_n)$ in the proposed interpretation in the text.

*Additional Reference*

Bhaskaran K, dos-Santos-Silva I, Leon DA, Douglas IJ, Smeeth L. Association of BMI with overall and cause-specific mortality: a population-based cohort study of 3• 6 million adults in the UK. The Lancet Diabetes & Endocrinology. 2018 Dec 1;6(12):944-53.

Trudel-Fitzgerald C, Millstein R, von Hippel C, Howe R, Tomasso LP, Wagner G, VanderWeele TJ (2019). Psychological well-being as part of the public health debate? Insight into dimensions, interventions, and policy. BMC Public Health, 15;9(12):e033697.